\documentstyle[aps,prl,epsf,twocolumn]{revtex}
\begin{document}
\draft
\title{{\rm Physical Review Letters}\hfill {\sl Version of \today}\\~~\\
Dynamics and Wrinkling of Radially Propagating Fronts Inferred from Scaling
Laws in Channel Geometries}
\author {Barak Galanti, Oleg Kupervasser, Zeev Olami and
Itamar  Procaccia$^*$ }
\address{Department of~~Chemical Physics,\\
 The Weizmann Institute of Science,
Rehovot 76100, Israel\\$^*$Address for fall 1997: The Rockefeller University,
1230 York Ave., New York, NY 10021}
\maketitle
\begin{abstract}
Flame Propagation is used as a prototypical example of expanding fronts
that wrinkle without limit in radial geometries but reach a simple
 shape in channel geometry. We show
that the relevant scaling laws that govern the radial growth can be
inferred once the
simpler channel geometry is understood in detail. In radial geometries (in
contrast to
channel geometries) the effect of external
noise is crucial in accelerating and wrinkling the fronts. Nevertheless,
once the interrelations between
system size, velocity of propagation and noise level are understood in channel
geometry, the scaling laws for radial growth follow.
\end{abstract}

The main idea of this Letter is that in order to derive scaling laws for
unstable front
propagation in radial geometry, it is useful to study noisy propagation in
channel geometries,
in which the noiseless dynamics results usually in simple shapes of the
advancing fronts \cite{Pel}.
Famous examples of such situations are Laplacian growth patterns and unstable
flame fronts. Laplacian growth in a channel geometry results in a single
finger whose width
is uniquely determined by the channel width $L$, the velocity of the tip
$v$ and the surface
tension\cite{Lap}. Laplacian growth in radial geometries results in convoluted
 and
complex structures whose
full characterization still eludes repeated theoretical
attacks\cite{Lap,BS,Vic,95HHZ}. Flame propagation offers a similar
situation; in channel geometries the flame front looks like a giant cusp
whose velocity
$v$ is constant\cite{85TFH}. In radial geometries the flame front accelerates
 with
time, and while
propagating it wrinkles by the addition of hierarchies of cusps of all
sizes\cite{89GIS,94FSF,89J,95J,KOP95,OGKP}.

The understanding of radial geometries requires control of the effect of
noise on the
unstable dynamics of propagation. It is particularly difficult to achieve
such a control
in radial geometries due to the vagueness of the distinction between
external noise and noisy
initial conditions. Channel geometries are simpler when they exhibit a
stable solution for
growth in the noiseless limit. One can then study the effects of external
noise in such
geometries without any ambiguity. If one finds rules to translate the
resulting understanding
of the effects of noise in channel growth to radial geometries, one can
derive the scaling laws in the
later situation in a satisfactory manner. We will exemplify the details of
such a translation
in the context of premixed flames that exist as self sustaining fronts of
exothermic chemical
reactions in gaseous combustion. But our contention is that similar ideas
should be fruitful
also in other contexts of unstable front propagation. Needless to say, there
are aspects of the front dynamics and statistics in the radial geometry
that {\em cannot} be explained from observations of fronts in a channel
geometry; examples of such aspects are discussed at the end of this
Letter.

Mathematically our example is described \cite{94FSF} by an equation of
motion for the angle-
dependent modulus of the  radius vector of the flame front,  $R(\theta,t)$:
\begin{eqnarray}
{\partial R \over \partial t}&=&
 {U_b\over 2{R_0}^2(t)}\left({\partial R \over \partial \theta }\right)^2
 +{D_M\over {R_0}^2(t)}{\partial^2 R\over \partial\theta^2}\\ \nonumber &+&
{\gamma
 U_b\over 2R_0 (t)} I(R)+U_b \ . \label{Eqdim}
\end{eqnarray}
Here 0$ <\theta <  2\pi$ is an angle and the constants $U_b,D_M$ and
$\gamma$ are the
front velocity for an ideal cylindrical front, the Markstein diffusivity
and the thermal
expansion coefficient respectively. $R_0(t)$ is the mean
radius of the propagating flame:
\begin{equation}
R_0 (t)={1\over 2\pi}\int_{0}^{2\pi}R(\theta,t)d\theta \  . \label{R0}
\end{equation}
The functional $I(R)$ is best represented in terms
of its Fourier decomposition. Its Fourier component is $|k|R_k$ where $R_k$
is the
Fourier component of $R$. Simulations of this equation, as well as
experiments in
the parameter regime for which this equation is purportedly relevant,
indicate that for large times $R_0$ grows as a power in time
\begin{equation}
R_0(t) = (const+t)^{\beta} \ , \label{accel}
\end{equation}
with $\beta>1$, and that the width of the interface $W$
grows with $R_0$ as
\begin{equation}
W(t) \sim R_0(t)^\chi \ , \label{scaling}
\end{equation}
with $\chi<1$. Note that this exponent
may differ significantly from the exponent which characterizes
the scaling of correlation functions \cite{95OZP}. This phenomenon is common to
cases in which large features
dominate the width of the graph. In a previous publication \cite{KOP95} we
showed that
$\beta$ and $\chi$
satisfy the scaling relation $\beta=1/\chi$. In this Letter we determine the
numerical value of these exponents from the analysis of noisy channel growth.

In channels there is a natural lengthscale, the width $\tilde L$ of the
channel. The
translation of channel results to radial geometry will be based on the
identification
in the latter context of the time dependent scale ${\cal L}(t)$ that plays
the role of
$\tilde L$ in the former. To do this we need first to review briefly the
main pertinent results
for noisy channel growth. In channel geometry the equation of motion is
written in terms
of the position $h(x,t)$ of the
flame front above the $x$-axis. After appropriate rescalings \cite{85TFH} it
reads:
\FL
\begin{equation}
{\partial h(x,t) \over \partial t}=\!
{1\over 2}\!\left[{\partial h(x,t) \over \partial x }\right]^2
 \!\!+\!\!\nu{\partial^2 h(x,t)\over \partial x^2}+ I\{h(x,t)\}\!+\!1.
\label{Eqnondim}
\end{equation}
It is convenient to rescale the domain size further to $0<\theta<2\pi$, and
to change
variables to $u(\theta,t)\equiv {\partial h(\theta,t)/\partial\theta}$.
In terms of this function we find
\FL
\begin{equation}
{\partial u(\theta,t) \over \partial t}\!=\!
{u(\theta,t)\over L^2}{\partial u(\theta,t) \over \partial \theta }
\!+\!\!{\nu\over L^2}{\partial^2 u(\theta,t)\over \partial \theta^2}\!+
\!\!{1\over
L}I\{u(\theta,t)\}
\label{equ}
\end{equation}
where $L=\tilde L/2\pi$. In noiseless conditions this equation admits exact
solutions
that are represented in terms of $N$ poles whose position $z_j(t)\equiv
x_j(t)+iy_j(t)$
in the complex plane is time dependent:
\begin{equation}
u(\theta,t)=\nu\sum_{j=1}^{N}\cot \left[{\theta-z_j(t) \over 2}\right]
   + c.c.\ , \label{upoles}
\end{equation}
The steady state for channel propagation is unique and linearly stable; it
consists of
$N(L)$ poles which are aligned on one line parallel to the imaginary axis. The
geometric appearance
of the flame front is a giant cusp, analogous to the single finger in the
case of
Laplacian growth in a channel. The height of the cusp is proportional to
$L$, and the
propagation velocity is a constant of the motion. The number of poles
in the giant cusp is linear in $L$,
\begin{equation}
N(L)=\Big[{1\over 2}\Big({L\over 2\pi}+1\Big)\Big] \ . \label{NofL}
\end{equation}

The introduction of additive random noise to the dynamics changes the
picture qualitatively. It is convenient to add noise to the equation of
motion in Fourier
representation by adding
a white noise $\eta_k$ for every $k$ mode. The noise correlation function
satisfies the relation
$<\eta_k(t)\eta_{k'}(t')>=\delta(k+k')\delta(t-t')2f/L$. The noise in our
simulations is taken from a flat distribution
in the interval $[-\sqrt{2f/L},\sqrt{2f/L}]$;  this guarantees
that when the system size changes, the typical noise per unit
length of the flame front remains constant. It was shown \cite{OGKP} that for moderate
but fixed noise levels
the average velocity $v$ of the front increases with $L$ as a power law.
In our present simulations we found
\begin{equation}
v\sim L^{\mu} \ , \quad \mu\approx 0.35\pm 0.03 \ . \label{vlScale}
\end{equation}
For a fixed system size $L$ the velocity has also a power law dependence on the
level of the noise, but with a much smaller exponent:
$v\sim f^\xi \ , \quad \xi\approx 0.02$.
These results were understood theoretically by analysing the noisy creation of
new poles that interact with the poles defining the giant cusp\cite{OGKP}. It
 was
shown that the system
is always linearly stable or marginally stable for the introduction of new
poles, but with
increasing $L$ all the eigenvalues (which are negative for a positive
measure of $L$ values)
all decrease in absolute magnitude like $1/L^2$, making the system more and
more susceptible to
noisy perturbations \cite{OGKP}. The picture used remains valid
as long as the poles that are introduced by the noisy perturbation do not
destroy the identity
of the giant cusp. Indeed, the numerical simulations show that in the
presence of moderate
noise the additional poles appear as smaller cusps that are constantly
running towards
the giant cusp. Our point here, is not to predict the numerical values of the
scaling exponents in the {\em channel} (this was done in ref.\cite{OGKP},
but to use them to predict the scaling exponents characterizing the
acceleration and the
geometry of the flame front in {\em radial} geometry.

Superficially it seems that in radial geometry the growth pattern is
qualitatively different.
In fact, close observation of the growth patterns (see Fig.1) shows that
most of the time
there exist some big cusps that attract other smaller cusps,
but that every
now and then ``new" big cusps form and begin to act as local absorbers of
small cusps
that appear randomly.  The understanding of this phenomenon gives the clue
how to
translate results from channels to radial growth.
 \begin{figure}
 \epsfxsize=6.5truecm
 \epsfbox{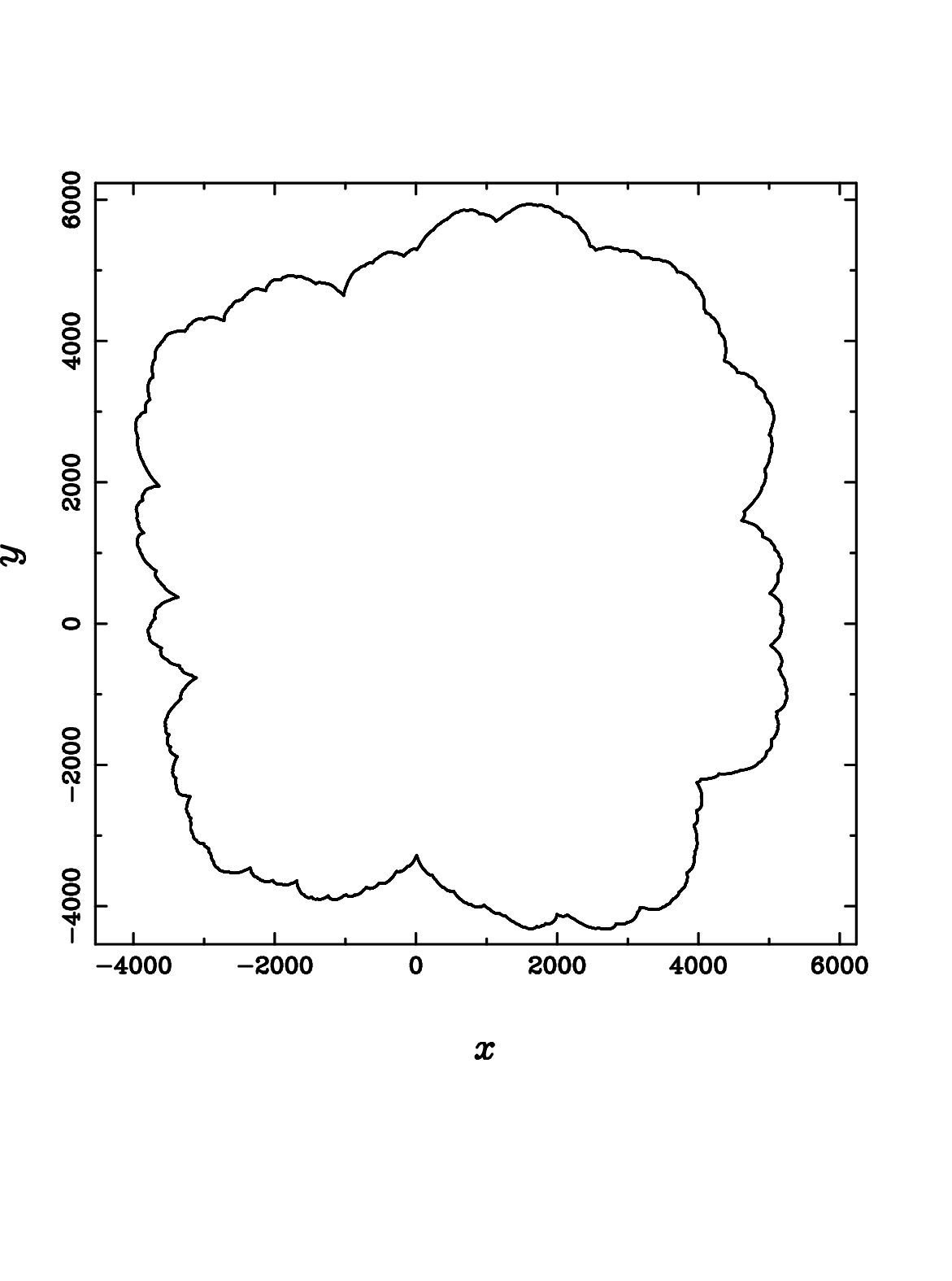}
 \caption{Simulations of the outward propagating flame front. Note that
there is a wide distribution of cusp sizes.}
\label{fig:fig1}
 \end{figure}

In the radial geometry we
non-dimensionalize
Equation (\ref{Eqdim}) by rescaling $R\to r,R_0\to r_0$ and $t\to \tau$,
and finally rewriting the equation in terms of
$u \equiv {\partial r/ \partial \theta}$:
\begin{equation}
{\partial u \over \partial \tau}={u \over r_{0}^2}
   {\partial u \over \partial \theta}+{1\over r_{0}^2}{\partial^2u
   \over \partial \theta^2}+{\gamma\over 2 r_0}I\{ u\} \ . \label{eqfinal}
\end{equation}
To complete this equation we need a second one for $r_0(t)$. This equation
is obtained
by averaging (\ref{Eqdim}) over the angles (i.e. operating on the equation
with ${1\over
2\pi}\int_{0}^{2\pi} d\theta$):
\begin{equation}
{dr_0 \over d\tau}={1  \over 2r_{0}^2}{1 \over
2\pi}\int_{0}^{2\pi}u^2d\theta +1 \ .
\label{eqr0}
\end{equation}
These equations again admit exact solutions in terms of poles, of the form
of Eq.(7).
It is easy write down the equations of motion of the poles and check that
the poles
are attractive along the real direction (which means physically that they are
attracted along the angular coordinate) but they are repulsive along the
imaginary direction,
which is associated with the radial coordinate. If it were not for the
stretching that is
caused by the increase of the radius (and with it the perimeter), all the
poles would have
coalesced into one giant cusp.
Thus we have a competition between pole attraction and stretching. Since the
attraction decreases
with the distance between the poles in the angular directions, there is
always an initial  critical
length scale above which poles cannot coalesce their real
coordinates when time progresses.

Suppose now that noise adds new poles to the system. The poles do
not necessarily
merge their real positions with existing cusps. If we have a large cusp
made from the merging
of the real coordinates $x_c$ of $N_c$ poles, we want to know whether a
nearby pole with real
coordinate $x_1$ will merge with this large cusp. The answer will depend of
course on the
distance $D\equiv r_0|x_c-x_1|$. A direct calculation \cite{KOP95}, using
the equation
of motions for the
poles shows that there exists a critical length ${\cal L}(r_0)$ such
that if $D>{\cal L}(r_0)$
the single pole never merges with the giant cusp. The result of the
calculation is that
\begin{equation}
{\cal L}\sim r_0^{1/\beta} \ . \label{Lbeta2}
\end{equation}
Note that a failure of a single pole to be attracted to a large cusp
means that tip-splitting has occurred. This is the exact analog of tip-splitting
in Laplacian growth.

It is now time to relate the channel and radial geometries. We identify the
typical scale in the radial geometry as ${\cal L}\sim W\sim r_0^\chi$. On the
one hand this leads to the scaling relation $\beta=1/\chi$. On the other
hand we use the result established in a channel, (\ref{vlScale}), with this
identification of a scale, and find
$\dot  r_0 = r_0 ^{\chi \mu}$. Comparing with (\ref{accel}) we find:
\begin{equation}
\beta = {1\over (1- \chi \mu)}\label{beta} \ . \label{betamu}
\end{equation}

This result leads us to expect two dynamical regimes for our problem.
Starting from
smooth initial conditions, in relatively
short times the roughness exponent remains close to unity.
This is mainly since the typical scale ${\cal L}$ is not relevant yet,
and most of the poles that are generated by noise
merge into a few larger cusps.
In later times the roughening exponent settles at its asymptotic value,
and all the asymptotic scaling relations used above become valid.
We thus expect $\beta$ to decrease from $1/(1-\mu)$ to an
asymptotic value determined by $\chi=1/\beta$ in (\ref{betamu}):
\begin{equation}
\beta=1+\mu \approx 1.35 \pm 0.03 \ . \label{predbeta}
\end{equation}
The expected value of $\chi$ is thus $\chi=0.74\pm 0.03$.
\begin{figure}
 \epsfxsize=6.5truecm
\epsfbox{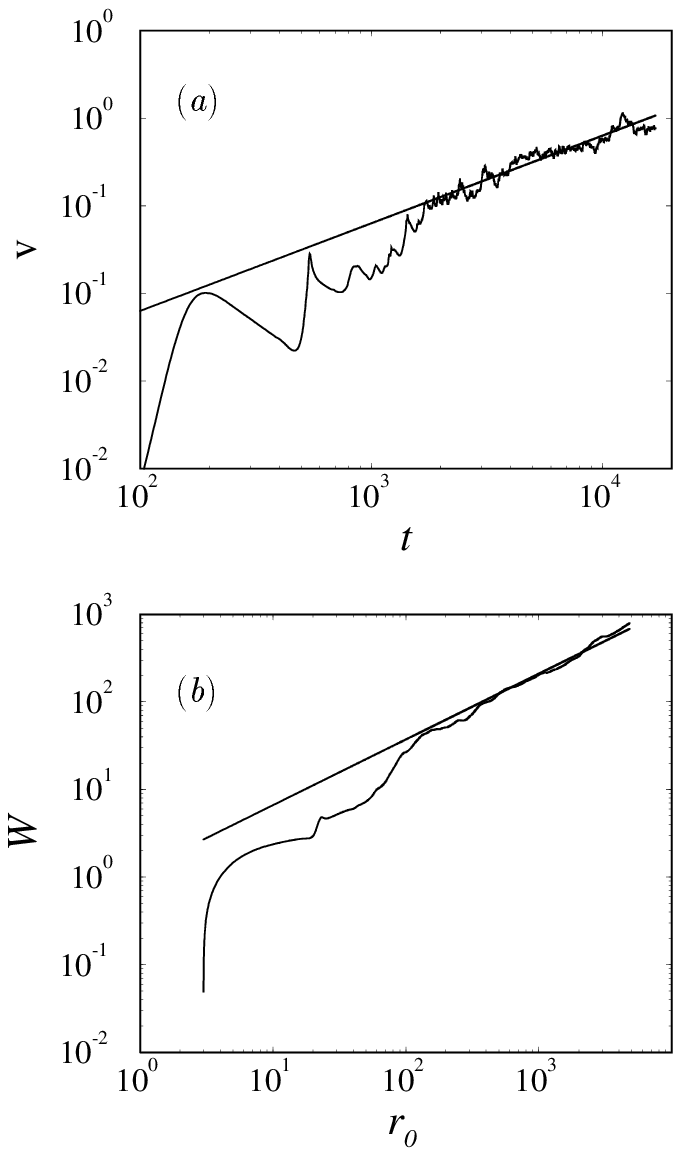}
\caption{Panel a: a logarithmic plot of the velocity versus time for a
radially evolving
 system. The parameters
of the simulation are: $f=10^{-8}$, $\gamma=0.8$, $\nu=1$. Panel
b:Logarithmic plot of
the width of the flame
front as a function of the mean radius.}
\label{fig:fig2}
\end{figure}
We tested these predictions in numerical
simulations. We integrated Eq.(\ref{equ}), and in
Fig. 2 we display the results
for the growth velocity as a function of time.  After a
limited domain of exponential growth we observe a continuous reduction
of the time dependent exponent.
In the initial region we get $\beta  = 1.65\pm 0.1$ while in the final decade
of the temporal range we find
$\beta  = 1.35 \pm 0.1$. We consider this a good agreement with
 (\ref{predbeta}).
 A second important test is provided by measuring the width of the system as
a function of the radius, see Fig.2b. Again we observe a cross-over related to
the initial dynamics; In the last temporal decade the exponent settles at
$\chi= 0.75\pm .1$.
We conclude that at times large enough to observe the asymptotics our
 predictions
are verified.

Finally, we stress some differences between radial and channel geometries.
Fronts in a channel exhibit mainly one giant cusp which is only marginally
disturbed by the small cusps that are introduced by noise. In the radial
geometry, as can be concluded from the discussion above, there exist at
any time cusps of all sizes from the smallest to the largest. This broad
distribution of cusps (and scales) must influence correlation function
in ways that differ qualitatively from correlation functions computed in
channel geometries.
\begin{figure}
 \epsfxsize=6.5truecm
\epsfbox{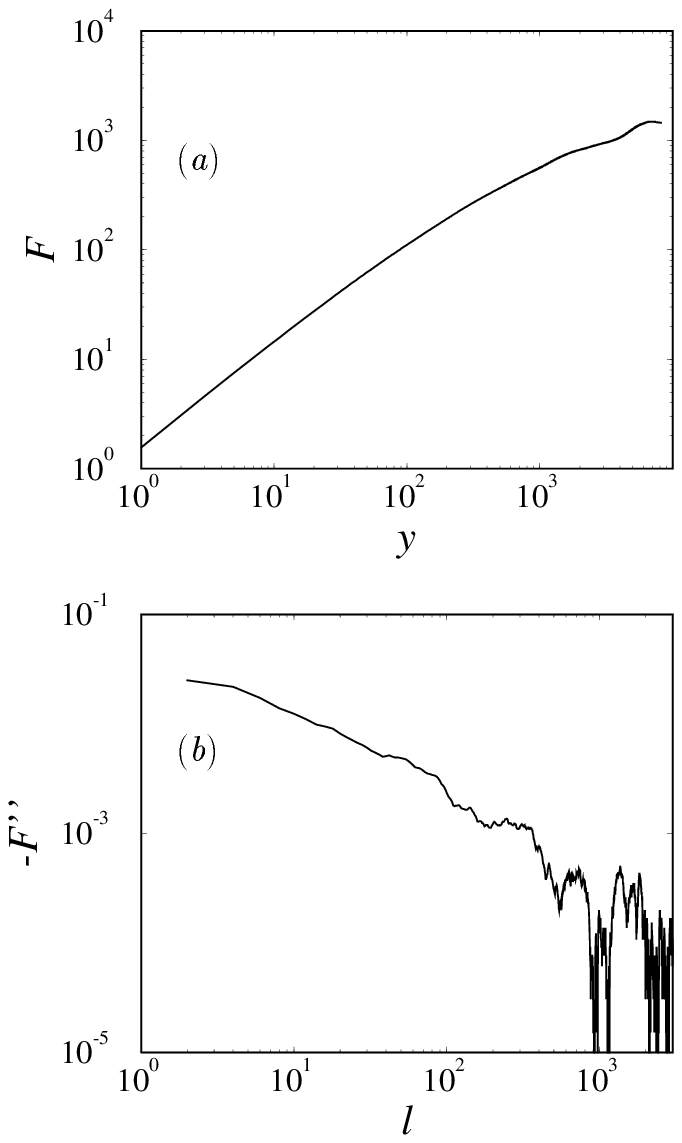}
\caption{Panel a: a logarithmic plot of the correlation function of the
interface.
Panel b: Second derivative of the correlation function of the interface.}
\end{figure}
To make the point clear we exhibit in Fig.3a the structure function
\begin{equation}
F(y)\equiv \sqrt{\left< |R(x+y)-R(x)|^2\right >} \label{F}
\end{equation}
computed for a typical radial front, with $x=R\theta$. To stress the
scaling region we exhibit
the second derivative of this function in Fig.3b. The low end of the
graph can be fitted well by a power law $y^{-\alpha}$ with $\alpha\approx 0.6$.
This indicates that $F(y)\approx Ay +By^{2-\alpha}$. In a channel
geometry we get entirely different structure functions that do not exhibit such
scaling functions at all. The way to understand this behavior in the radial
geometry is to consider a distribution of cusps that remain distinct from
each other but whose scales are distributed according to some distribution
$P(\ell)$.
For each of these cusps there is a contribution to the correlation function
of the form $f(y,\ell)\approx \ell g(y/\ell)$ where $g(x)$ is a scaling
function,
$g(x)\approx x$ for $x<1$ and $g(x)\approx$ constant for $x>1$. The total
correlation
can be estimated (when the poles are distinct) as
\begin{equation}
F(y)\approx \sum_\ell P(\ell)\ell g(y/\ell) \ .
\end{equation}
The first derivative leaves us with $\sum_\ell P(\ell)g'(y/\ell)$, and using
the fact that $g'$ vanishes for $x>1$ we estimate $F'(y)=\sum_{\ell=y}^W
P(\ell)$.
The second derivative yields $F''(y)\approx - P(y)$. Thus the structure
function is
determined by the scale distribution of cusps, and if the latter is a power law,
this should be seen in the second derivative of $F(y)$ as demonstrated in
Fig.3.
The conclusion of this analysis is that the radial case exhibits a scaling
function that characterizes the distributions of cusps, $P(\ell)\approx
\ell^{-\alpha}$.

In summary, we demonstrated that it is possible to use information about
noisy channel dynamics
to predict nontrivial features of the radial evolution, such as the
acceleration and roughening exponents. It would be worthwhile to examine
similar ideas in the context of Laplacian growth patterns.
\acknowledgments This work has been supported in part by the Basic Research
Fund administered by the Israel Academy for Science and Humanities. IP
thanks Mitchell Feigenbaum for the hospitality at the Rockefeller University.

\end{document}